\documentclass[%
prl,
 jmp,%
 amsmath,amssymb,
 reprint,%
showkeys,twocolumn]{revtex4-1}

\usepackage{graphicx}
\usepackage{dcolumn}
\usepackage{bm}

\begin{document}

\title{Temperature dependence of slow positron reemission from metals}
\date{November 14, 2014}

\author{Reza Rahemi}
\email{rahemi@me.com}


\begin{abstract}
Intensity of positron reemission from a metal is related to its work function. This property is dependent on temperature and can be modified with temperature control. In this article a simple model is proposed to explain and predict the variation of the positron work function with temperature. The dependence of slow positron yield of metals on temperature is predicted based on this model and the critical temperatures at which the slow positron yield is maximum is calculated. The proposed model is consistent with experimental observations.
\end{abstract}

            
                                                   
\maketitle


\section{Introduction}

Positron is often used in many applications and investigation methods, e.g, Positron Emission Tomography (PET), Positronium Annihilation Lifetime Spectroscopy (PALS) and positron reemission microscopy (PRM).

Positron Emission Tomography (PET) is a medical imaging technique which uses positrons from the decay of a radio-isotope which is inserted into the body by the means of injection or inhalation of radio-pharmaceuticals. Decaying positrons have energies greater than their rest mass. When positrons interact with the nearby electrons, they annihilate by emitting photons which are detectable. This results in production of high resolution images of the tissues \cite{ollinger1997positron, bailey2005positron}. 

Another important application is the use of the positron beam as a probe to analyze defects in materials, since around a vacancy-type defect, the electron density is locally reduced which results in positron lifetime being longer than that in defect-free lattice \citep{Tuomisto}. This method is often used to investigate the structures of defects in semiconductors, polymers and nano-structured materials and also to study temperature induced phase transitions in High $T_c$ superconductors \cite{jean1990positron, ferrell1956theory, krause1999positron, zubiaga2007positron, haldar1996temperature, bas2004positron, huang1990positron, peter1998positron, zhang2002study, wiktor2014positron, teng1987positron, dong2015application, fan2015defect, shpotyuk2015free}.

Furthermore, positron reemission microscopes (PRM) are used for surface studies in biological tissues and semiconductors \cite{House1988,brandes1988submicron,brandes1991positron, frieze1990positron,hulett1984generation,PhysRevLett.61.492, PhysRevB.43.10103,goodyear1995development, canter1995contrast}. Understanding the temperature dependence of the spontaneous reemission of positrons from the metal surfaces can guide selecting or designing better materials for relevant applications.     

Positron reemission intensity is related to the work function of positrons which is dependent on temperature. However, the effect of temperature on the position work function is not well understood. It is possible to investigate this effect by looking at the dependence of the work function of its anti-particle, electron, on temperature. The electron work function is a barrier for the electrons to be moved from inside a solid to a point in vacuum right outside the solid surface \citep{ashcroft1981solid}. As the temperature increases, this barrier decreases, since electrons are thermally excited and easier to escape from the metal surface. The effect of temperature on electron work function was previously investigated by the present author and a T-dependent electron work function model has been proposed \citep{rahemi2015variation}. 

The objective of this study is to establish a temperature dependent work function relationship for positron, based on the previous study regarding that of the electron work function and the existing knowledge regarding the temperature dependence of the positronium atom. 

\section{Variation in positron work function with temperature}

According to Lang and Kohen \citep{Lang}, the electron work function consists of two separate contributions. One is the bulk electron chemical potential, $\mu^-$ which is the difference in the energy of ground state before and after the removal of an electron from the solid. The other is the surface dipole moment, $\Delta$, which is the potential caused by electrons spilling out beyond the metal surface.
\begin{equation} \label{Eq_1_} 
\varphi^-=-\mu^{-}+\Delta
\end{equation} 
and

\begin{equation} \label{Eq_2_} 
\varphi^+=-\mu^+-\Delta 
\end{equation} 
where $\varphi^+$ and $\varphi^-$ are work functions of positron and electron, and $\mu^+$ and $\mu^-$ are their chemical potentials, respectively. 

Positronium is a short-lived hydrogen-like atom composed of an electron and a positron (the antiparticle of the electron). It decays by annihilation to produce two or more (less often) photons. This exotic atom has a thermalization time of approximately $10^{-12}$s in condensed matter \citep{Berko}. Since the positronium work function is related to the electron work function, varies with temperature, and is a general property of metals rather than a characteristic of a particular metal \citep{Rosenberg}, it can be used as a probe to investigate the variation in the positron work function of metals with temperature.

The positronium work function, composed of electron work function ($\varphi^-$) and positron work function ($\varphi^+$), is expressed as

\begin{equation} \label{Eq_3_} 
\varphi^{Ps}=\varphi^++\varphi^--6.8eV 
\end{equation} 
where 6.8 eV is the binding energy of the positronium atom \citep{Rosenberg,Nieminen,Panda}. Differentiating equation \ref{Eq_3_} with respect to temperature gives

\begin{equation} \label{Eq_4_} 
\frac{d\varphi^{Ps}}{dT}=\frac{d\varphi^+}{dT}+\frac{d\varphi^-}{dT}
\end{equation} 

A temperature-dependent work function has been derived via generalization of Lennard-Jones potential and is expressed as \citep{rahemi2015variation}: 

\begin{equation}\label{Eq_5_}
\varphi^-=\varphi_0^--\gamma\frac{(k_BT)^2}{\varphi_0^-}
\end{equation}
where $\varphi_0^-$ is the electron work function of a material at $T=0 \ K$ and $\gamma $ is a calculable material coefficient. More details about calculation of $\gamma $ is given in ref. \citep{rahemi2015variation} and some values are presented in Table \ref{tab_1}.

Differentiating eq. (\ref{Eq_5_}), yields

\begin{equation} \label{Eq_6_} 
\frac{d\varphi^-}{dT}=-2\gamma\frac{k_B^2}{\varphi_0^-}T
\end{equation} 
Combining equations \ref{Eq_4_} and \ref{Eq_6_} we have:

\begin{equation} \label{Eq_7_} 
\frac{d\varphi^+}{dT}=(\frac{d\varphi^{Ps}}{dT}+\frac{2\gamma}{\varphi_0^-}k_B^2 T)
\end{equation} 
Integrating equation \ref{Eq_7_} yields

\begin{equation} \label{Eq_8_} 
\varphi^+=\int_{0}^{T}(\frac{d\varphi^{Ps}}{dT}+\frac{2\gamma}{\varphi_0^-}k_B^2T')dT'= \\
\frac{d\varphi^{Ps}}{dT}T+\frac{\gamma}{\varphi_0^-}(k_BT)^2 
\end{equation} 

Experimental results of reemission of implanted slow positron (with energies in the order of eV to KeV) suggest that as the temperature rises from 50 $K$ to about 200 $K$, the positron yield should also rise \cite{Mills1978,Huttunen1990,Schultz} The positron yield ($Y_{Ps}$) is related to the positron work function \citep{Schultz}. This can be realized as a decrease in the work function of positron and subsequently an increase in the positron reemission intensity given that the positrons with work functions equal to that of a metal are re-emitted from the surface \cite{sudarshan2013positron}. Assuming that this relationship is linear, the positron yield may be written as

\begin{equation} \label{Eq_9_} 
Y_{ps}=C (\frac{d\varphi^{Ps}}{dT}T+\frac{\gamma}{\varphi_0^-}(k_BT)^2)+y_0
\end{equation} 
where $c$ is a constant of proportionality (negative) and $y_{0}$ is the initial value for yield at the absolute zero. 

Experimental results from \citet{Schultz} and predicted values using eq. (\ref{Eq_9_}) are plotted and presented in Fig. 1. Values of $c$ and $y_{0}$ are obtained by fitting the experimental data and the average value for $\frac{d\varphi^{Ps}}{ dT}$ is used from available experimental measurements. $\frac{d\varphi^{Ps}}{ dT}$ is a constant and a general property of metals, equal to $-(6 \pm 3) k_B$ \citep{Rosenberg}. The experimental and theoretical curves are in good agreement.  

\begin{figure}
\centering
\scalebox{0.5}{\includegraphics{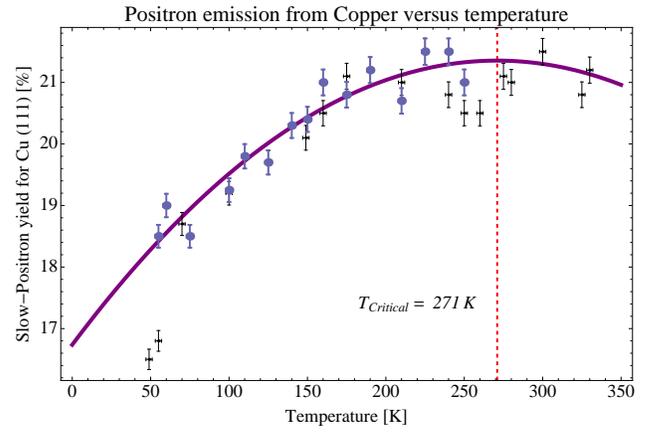}}
\caption{Variations in slow positron yield with temperature. The data points are from ref. \citep{Schultz} and the solid line is predicted based on eq. \ref{Eq_9_}. }
\end{figure}

\section{More information from the model}

It is evident that the slow positron yield increases as the temperature rises. However, after reaching a critical temperature, the yield starts to decrease with continuously increasing temperature. Such phenomenon can be explained based on the energy loss of positron at surface. The proposed model also agrees with this observation. As eq. (\ref{Eq_2_}) suggests, the positron work function is contributed by its bulk chemical potential and surface dipole moment. Differentiating eq. (\ref{Eq_2_}) with respect to the temperature gives

\begin{equation} \label{Eq_10_} 
\frac{d\varphi^+}{dT}=-\frac{d\mu^+}{dT}-\frac{d\Delta}{dT}
\end{equation} 

Value of $\frac{d\mu^+}{dT}$ is negative, since the thermal positrons lower their energies by adjusting to the ionic density fluctuations \citep{Boev}. The positron chemical potential has three major contributors: zero-point energy due to the motion of the positron in the ionic lattice, the positron-electron correlations and the positron-phonon coupling, which is negligible compared to the other two \citep{Rosenberg}.

The term $\frac{d\Delta}{dT}$ is positive, since as more electrons are excited thermally, the attraction of positrons into the vacuum by the electrostatic dipole potential increases \citep{Panda}. Although the dipole term is small compared to the bulk potential terms for low-density metals \citep{Lang}, there is no priori assumption that their variation with temperature should also be smaller. At this point, it is understood that at a certain temperature (i.e the critical temperature), the sum of the negative $\frac{d\mu^+}{dT}$ and the positive $\frac{d\Delta}{dT}$  becomes zero. This critical temperature can be calculated by setting eq.\ref{Eq_7_} equal to zero. Then $T_{\textit{Critical}}$ becomes:

\begin{equation} \label{Eq_11_} 
T_{\textit{Critical}}=-\frac{1}{2}\frac{d\varphi^{Ps}}{dT}\frac{\varphi_0^-}{\gamma k_B^2}
\end{equation}

This critical temperature is of significance, since the intensity or yield of positrons is maximum at this temperature. Using this critical temperature as a guide, one may control and maximize positron beams for optimal applications.

\begin{table}[t] 

\begin{centering}

\begin{ruledtabular}

\begin{tabular}{cccccc}

Metal&
$\gamma$&
$\varphi^{-}_{295 K}[eV]$&
$\varphi^{-}_{0} [eV]$&
$T_{C} (Cal.) [K]$&
$T_{C} (Exp.) [K]$

\tabularnewline
\hline 
\tabularnewline
Al&
583&
4.28&
4.37&
$261 \pm 130$&
$273 \pm 150$ 

\tabularnewline
Ag&
478&
4.26&
4.33&
$315 \pm 157$&
$200 \pm 100$ 

\tabularnewline
Cu&
307&
4.5&
4.54&
$515 \pm 257$&
$300 \pm 50$ 
\end{tabular}
\end{ruledtabular}
\caption{\label{tab_1}
Calculated values for the critical temperature, $T_{C} (Cal.) $, at which the positron yield becomes maximum, using equation (\ref{Eq_11_}). The experimental values, $T_{C} (Exp.)$, for alumnium, silver and copper are from experimental results reported by Ref. \cite{Mills1978}, \cite{Huttunen1990} and \cite{Schultz} respectively. $\gamma$ values are reported from Ref. \cite{rahemi2015variation} and $\varphi^{-}_{0} [eV]$ are calculated based on equation (\ref{Eq_5_}) and using available room temperature values for electron work function.}
\par\end{centering}
\end{table}

\section{Remarks}

A simple model to describe the temperature dependence of positron work function in metals is proposed. Based on this model, the temperature dependence of slow positron yield is predicted. The existence of a critical temperature at which the positron yield and consequently the positron beam intensity reach maxima is theoretically revealed, which was experimentally observed previously without quantitative explanation. This critical temperature is expressed as $T_{\textit{Critical}}=-\frac{1}{2}\frac{d\varphi^{Ps}}{dT}\frac{\varphi_0^-}{\gamma k_B^2}$ where $\gamma$ is a calculable material property which is dependent on the crystal structure. This model would enhance our understanding of the positron behaviour at different temperatures. Since the work function of the positrons is related to their work function, modification of this property will facilitate adjustment of their emission intensity and help tune the positron beams for optimal applications, via temperature control. The consistency between the prediction and experimental observations verifies the model and the relationship between positron yield and positron work function.

\acknowledgments

The author is grateful for financial support from the  Natural Sciences and Engineering Research Council of Canada (NSERC) and to Prof. Dongyang Li for taking time to review the introduction. 

\bibliography{references}

\end{document}